\documentclass[a4paper,11pt]{article}
\pdfoutput=1 
\usepackage{jheppub} 
\usepackage[T1]{fontenc} 
\makeatletter
\def\@fpheader{\relax}
\makeatother
\newcommand{\mdm}{M_\mathrm{DM}}
\newcommand{\sigmav}{\sigma v}
\newcommand{\sigmap}{\sigma_p}
\newcommand{\sigmaSI}{\sigma_p^{SI}}
\newcommand{\sigmaSD}{\sigma_p^{SD}}

\title{\boldmath On the Role of Neutrinos Telescopes in the Search for Dark Matter Annihilations in the Sun}

\author[a,b]{Nicolao Fornengo,}
\author[c,d]{Antonio Masiero,}
\author[e,f]{Farinaldo S. Queiroz,}
\author[g]{Carlos E. Yaguna}
\affiliation[a]{Dipartimento di Fisica, Universit\'a di Torino\\via P. Giuria,
1, I-10125 Torino, Italy}
\affiliation[b]{Istituto Nazionale di Fisica Nucleare, Sezione di Torino\\ via
P. Giuria, 1, I-10125 Torino, Italy}
\affiliation[c]{Dipartimento di Fisica e Astronomia  ``G. Galilei'', Universita di Padova, Italy}
\affiliation[d]{Istituto Nazionale Fisica Nucleare, Sezione di Padova, I-35131 Padova, Italy}
\affiliation[e]{Max-Planck-Institut f\"ur Kernphysik\\ Saupfercheckweg 1, 69117 Heidelberg, Germany}
\affiliation[f]{International Institute of Physics, Federal University of Rio Grande do Norte, Campus Universit\'ario, Lagoa Nova, Natal-RN 59078-970, Brazil}  
\affiliation[g]{Escuela de F\'{i}sica, Universidad Pedag\'ogica y Tecnol\'ogica de Colombia (UPTC)\\
Avenida Central del Norte, Tunja, Colombia}

\emailAdd{fornengo@to.infn.it} 
\emailAdd{masiero@pd.infn.it}
\emailAdd{queiroz@mpi-hd.mpg.de}
\emailAdd{carlos.yaguna@uptc.edu.co}

\abstract{The observation of GeV neutrinos coming from the Sun would be an unmistakable signal of dark matter. Current neutrino detectors have so far failed to detect such a signal, however,  and bounds from direct and indirect dark matter searches  may significantly restrict the possibility of observing it in future experiments such as Hyper-Kamiokande or IceCube-Gen2. In this work we assess, in the light of current data and of expected experimental sensitivities, the prospects for the detection of a neutrino signal from dark matter annihilations in the Sun.  To be as general as possible, equilibrium between the capture and the annihilation rates in the Sun is not assumed in our analysis; instead, the dark matter scattering and annihilation cross sections are taken as free and independent parameters. We consider capture via both spin-dependent and spin-independent interactions, and annihilations into three representative final states: $b\bar b$, $W^+W^-$, and $\tau^+\tau^-$. We find that when the  capture in the Sun is dominated by  spin-independent interactions, current direct detection bounds already preclude the observation of a neutrino signal in future experiments.  For capture via spin-dependent interactions,  a strong complementarity is observed, over most of the parameter space, between future neutrino detectors and planned direct and indirect dark matter detection experiments, such as PICO-500 and CTA. In this case, we also identify some regions of the parameter space that can be probed, via the neutrino flux from the Sun, only by future neutrino experiments.}

\begin{document} 
\maketitle
\flushbottom

\section{Introduction}
\label{sec:intro}

The identification of the dark matter, that mysterious form of matter that accounts for more than $25\%$ of the energy-density of the Universe, is one of the most important open problems in particle and astroparticle physics today \cite{Bertone:2004pz,Bertone:2016nfn,Queiroz:2016sxf,Kavanagh:2017hcl}. The first step towards its solution would likely be the detection of the  dark matter particle via  non-gravitational interactions.  That is the motivation behind the huge and varied experimental and theoretical effort currently underway aiming at observing a robust signal of the dark matter particle directly or indirectly \cite{Pato:2010zk,Cirelli:2012tf,Cremonesi:2013bma,Strigari:2013iaa,Iocco:2015xga,Conrad:2015bsa,Mayet:2016zxu,Roszkowski:2017nbc,Battaglieri:2017aum,Balazs:2017hxh,Arcadi:2017kky,Acharya:2017ttl}.

Direct detection experiments, on the one hand, search for the scattering of a dark matter particle off a target nucleus \cite{Angle:2011th,Aprile:2012nq,Abe:2014zcd,Angloher:2015ewa,Aprile:2016swn,Cui:2017nnn,Akerib:2017kat,Agnese:2017njq,Aprile:2017yea}, among other signatures \cite{Aguilar-Arevalo:2016zop,Essig:2017kqs,Cavoto:2017otc,Davis:2017noy}. These experiments have, over the past few years, significantly improved the constraints on the dark matter-proton elastic scattering cross section, $\sigmap$ \cite{Undagoitia:2015gya}.  It is useful to distinguish, according to the type of interaction,  two different  scattering cross sections: the spin-independent,  $\sigmaSI$,  and the spin-dependent, $\sigmaSD$.   The most stringent limits on $\sigmaSI$ are currently set by  the XENON1T and PANDA-X experiments \cite{Aprile:2017iyp,Cui:2017nnn}. Both are still running and are expected to continue improving their limits over the next two to three years. Even bigger  experiments with greater sensitivities, including XENON-nT \cite{Aprile:2015uzo}, Darwin \cite{Aalbers:2016jon}, DEAP-50T \cite{Fatemighomi:2016ree} and EURECA phase 2 \cite{Angloher:2014bua}, are  planned further into the future. On the spin-dependent front, the best limits on $\sigmaSD$ are set by the PICO-60 experiment \cite{Amole:2017dex}. The PICO collaboration \cite{picoproject} has also  proposed a 1-ton detector, PICO 500 \cite{pico500A,pico500}, that will increase the sensitivity to a spin-dependent interaction by more than one order of magnitude. 

Indirect detection experiments, on the other hand, try to detect the fluxes of SM particles (photons, neutrinos, $e^+$, $\bar p$) that are produced when two dark matter particles annihilate with each other \cite{Colafrancesco:2006he,Abramowski:2011hc,Hooper:2012sr,Bringmann:2012ez,Baratella:2013fya,Giesen:2015ufa,Gonzalez-Morales:2017jkx}. Such annihilation processes are more likely to occur in places where the dark matter density is high, such as the galactic center (GC), dwarf spheroidal galaxies (dSphs), or inside the Sun. Currently, the most stringent and solid bounds on the dark matter annihilation rate, $\sigmav$, stem from the gamma-ray fluxes observed by FERMI (dSphs) \cite{Ackermann:2015zua,Drlica-Wagner:2015xua} and H.E.S.S. (GC) \cite{Abdallah:2016ygi}. Such limits are usually displayed on the plane ($\mdm$, $\sigmav$) for given  final states (e.g. $W^+W^-$). In the near future, CTA, the Cherenkov Telescope Array, is expected to improve those limits in a significant way \cite{Acharya:2013sxa,Silverwood:2014yza,Hutten:2016jko,Roszkowski:2016bhs,Conrad:2016jww,Morselli:2017ree}.

Dark matter particles can also be captured by the  Sun and annihilate in its interior, producing a neutrino flux that can be observed at Earth (only the neutrinos can escape). If detected, such GeV neutrinos would provide unmistakable evidence in favor of (WIMP) dark matter \cite{Choi:2015ara,Aartsen:2016zhm,ElAisati:2017ppn,Ardid:2017lry} \footnote{We left out the discussion concerning neutrinos signals from other regions such as the galactic halo because in this case the possibility for a neutrino signal is hampered by the existence of more promising indirect detection probes \cite{Queiroz:2016zwd}. }.  The resulting neutrino flux depends, on the particle physics side, both on the dark matter scattering cross section off nuclei, which determines the capture rate in the Sun,  and on the dark matter annihilation rate ($\sigmav$) and final states, which dictate the efficiency of the neutrino production. Therefore, the neutrino signal from dark matter annihilation in the Sun, which has led to a multitude of studies \cite{Esmaili:2009ks,Taoso:2010tg,Fukushima:2012sp,Kundu:2011ek,Gao:2011bq,Ibarra:2013eba,Choi:2013eda,Chen:2014oaa,Berger:2014sqa,Blumenthal:2014cwa,Catena:2015uha,Kouvaris:2015nsa,Rott:2016mzs,Feng:2016ijc,Ardid:2017lry},  lies in-between direct and indirect detection. When equilibrium is reached between the capture and the annihilation rates in the Sun, the neutrino flux no longer depends on  $\sigmav$ but only  on the scattering cross section and the annihilation final states --and $\mdm$. This equilibrium condition, which is often assumed in the literature, simplifies the analysis and allows to translate the  experimental limits on the neutrino flux from the Sun on a bound on either $\sigmaSI$ or $\sigmaSD$ for given final states, typicallly $b\bar b$, $W^+W^-$ and $\tau^+\tau^-$.  Currently, the most stringent bounds come from Super-Kamiokande, at low dark matter masses ($\mdm\lesssim 200~$GeV), and from  IceCube. In the future, improved sensitivities to a neutrino flux from dark matter annihilations in the Sun  are expected from Hyper-Kamiokande (HK) \cite{Abe:2011ts} and IceCube-Gen2 (IC-Gen2) \cite{Aartsen:2014njl} as well as with KM3NeT \cite{Katz:2006wv,Adrian-Martinez:2016fdl}.  

Our goal in this paper is to assess the role of neutrino telescopes in the discovery of dark matter annihilations in the Sun in prospect with ongoing and future direct and indirect experiments. We start reviewing, in the next section, the key ingredients for a neutrino signal from dark matter annihilations in the Sun. Then, we outline the current and projected experimental sensitivities on such signals, and next present our findings exploiting the interplay with flagship experiments in the search for the direct and indirect detection of dark matter. Finally, we discuss our results and draw our conclusions.

\section{The neutrino flux from DM annihilations in the Sun}
Dark matter particles can accumulate in the core of the Sun and annihilate with one another, giving rise to a neutrino flux that could be detected at Earth. The theory behind these processes is by now well-established and the calculations necessary for a  reliable prediction of the neutrino flux have already been incorporated into multiple programs such as DarkSUSY \cite{Gondolo:2004sc}, micrOMEGAs \cite{Belanger:2013oya} or PPPC \cite{Baratella:2013fya}. For completeness, we will simply review here the most salient features.

%We use DarkSUSY \cite{Gondolo:2004sc} to compute the expected neutrino flux....

The evolution of the number of dark matter particles in the Sun, $N_\chi$,  is described by
\begin{equation}
\dot N_\chi=C_\chi-A_{\chi\chi}N_\chi^2
\end{equation}
where $C_\chi$ and $A_{\chi\chi}$ are respectively the capture and annihilation rates. The evaporation rate was neglected in this equation as it is relevant only for dark matter masses below few GeVs. The annihilation rate, $A_{\chi\chi}$, is determined by the dark matter annihilation cross section, 
\begin{equation}
A_{\chi\chi}=\frac{\langle\sigma v\rangle}{V_{eff}},
\end{equation}
where $V_{eff}$ is the effective volume of dark matter in the Sun \cite{Belanger:2013oya}. The capture rate, $C_\chi$, is instead set by the dark matter scattering cross section on nuclei. Its value for the Sun can be approximated as \cite{Baratella:2013fya}
\begin{equation}
C_\chi\approx 5.9\times 10^{26}\,\mathrm{s^{-1}}\left(\frac{\rho_{DM}}{0.3\frac{\mathrm{GeV}}{\mathrm{cm^3}}}\right)\left(\frac{100~\mathrm{GeV}}{\mdm}\right)^2\left(\frac{270\,\mathrm{km\,/s}}{v_0^{eff}}\right)^3\frac{\sigma_{SD}+1200\,\sigma_{SI}}{\mathrm{pb}}
\end{equation}

From this equation we can see that if the spin-dependent cross section ($\sigma_{SD}$) and the spin-independent one ($\sigma_{SI}$) had the same value, dark matter capture in the Sun would be largely dominated by spin-independent interactions. It turns out, though, that the limits on $\sigma_{SI}$ are orders of magnitude more stringent than those on $\sigma_{SD}$, which tends to favor capture via spin-dependent interactions. In our analysis, we will in any case consider separately capture via spin-independent interactions (in section \ref{sub:si}) and spin-dependent ones (in section \ref{sub:sd}). 

For the Sun, equilibrium between the capture and annihilation processes is reached when $\sqrt{C_\chi A_{\chi\chi}}t_\odot \gg 1$ with $t_\odot=4.6\times 10^9$ years.
In that case, which is often assumed, the annihilation rate is determined by the capture rate, $A_{\chi\chi}N_\chi^2=C_\chi$, which in turn depends on the dark matter scattering cross section off nuclei.  One of the novelties of our analysis is that we do not assume equilibrium.  Instead, we consider the dark matter scattering and annihilation cross sections as free parameters and determine the regions where equilibrium is reached. The dotted lines in figures \ref{fig:sibb}-\ref{fig:sdtautau} correspond to  the condition $\sqrt{C_\chi A_{\chi\chi}}t_\odot = 1$.

If the dark matter particle annihilates into the final state $f\bar f$ (being $f$ any SM particle), the neutrino flux at the Earth is given by
\begin{equation}
\frac{d\phi_\nu}{dE_\nu}=\frac{1}{4\pi d_\odot^2}\frac 12 A_{\chi\chi}N_\chi^2 \frac{dN_f}{dE},
\end{equation}
where $N_f$ is the neutrino spectrum from that final state. Such spectra were calculated, taking into account oscillations and medium effects, in \cite{Cirelli:2005gh}. In our analysis we consider three possible final states: $b\bar b$, $W^+W^-$, and $\tau^+\tau^-$. The first one gives rise to a \emph{soft} neutrino spectrum whereas the other two yield  \emph{hard} spectra, offering better prospects for  detection. 

\section{Current limits and future sensitivities}
\label{sec:current}
As seen in  the previous section, the neutrino signal from dark matter annihilations in the Sun depends, on the particle physics side, on just four quantities: the dark matter mass ($\mdm$), the dark matter scattering cross section off protons ($\sigma_p$), the dark matter annihilation branching fractions into different final states, and the dark matter total annihilation rate ($\sigma v$). Different combinations of these quantities are constrained by neutrino experiments and also by direct and indirect detection experiments. In this section we briefly review these constraints and the expected future sensitivities. 

Neutrino detectors could directly observe the GeV neutrinos produced by  the annihilation of dark matter particles captured in the Sun. So far, no evidence of such  neutrinos has been detected, implying a limit on its flux. During the last few years, data from Baksan \cite{Boliev:2013ai}, Super-Kamiokande \cite{Choi:2015ara}, ANTARES \cite{Adrian-Martinez:2016gti}, and IceCube \cite{Aartsen:2016zhm}  has been used to set  significant limits on the neutrino flux.  

At low dark matter masses, $\mdm\lesssim 100$ GeV, the most stringent constraints come from Super-Kamiokande (SK). They are based on 3903 days of data and were translated, assuming equilibrium, into upper bounds on the spin-dependent or spin-independent cross sections for a given dark matter mass (below 200 GeV) and annihilation final state --see Table I of \cite{Choi:2015ara}.   At higher dark matter masses, it is IceCube (IC) that sets the most stringent bounds on the neutrino flux from the Sun \cite{Aartsen:2016zhm}. The latest IC limits were based on 3 years of data and covered the mass range between $20$ GeV and $10$ TeV. As before, they can be used to set constraints, assuming equilibrium, on the spin-dependent and the spin-independent dark matter scattering  cross sections.  

The standard way of constraining these scattering cross sections is, however, via dark matter direct detection experiments \cite{Undagoitia:2015gya}.  This year, new and more stringent limits on the scattering cross section were released for both spin-dependent and spin-independent interactions.  For spin-independent interactions, the most stringent limits were obtained by the XENON1T collaboration \cite{Aprile:2017iyp} (recently the PANDAX collaboration announced slightly stronger limits \cite{Cui:2017nnn}). They are based on 34.2 lives day of data and supersede the previous bounds from LUX \cite{Akerib:2016vxi}. For spin-dependent interactions, the best constraints on the scattering cross section off protons  was set by the PICO-60 collaboration \cite{Amole:2017dex}. These limits improved previous results by more than one order of magnitude. One of the goals of this work is precisely to analyze the impact that these new  limits  have on the possible observation of a neutrino signal from dark matter annihilation in the Sun.

Regarding future neutrino detectors, we consider Hyper-Kamiokande \cite{Abe:2011ts} and IceCube-Gen2 \cite{Aartsen:2014njl}. Hyper-Kamiokande (HK),  a successor of Super-Kamiokande (SK), is a next generation water Cherenkov detector to be built in Japan. With a fiducial mass about  20 times larger than that of SK, HK  is expected to improve the currents bounds on proton decay by about an order of magnitude and to play a crucial role in answering some of the most important open questions in neutrino physics \cite{Abe:2015zbg,Yano:2016rkf,Yokoyama:2017mnt}.  Regarding dark matter, the sensitivity of HK to a neutrino signal from dark matter annihilation in the Sun is expected to be a factor 3 or 4 greater than that of SK \cite{Labarga:2017rfo}.  IceCube-Gen2, currently under design study, is a next generation Antartic neutrino observatory that aims at increasing the instrumented volume to about $10\,\mathrm{km}^3$ \cite{Aartsen:2014njl}. Current proposals add 120 strings to the detector, with different spacings and geometries \cite{Blaufuss:2015muc,Meagher:2017zvi}. Since the design details still need to be finalized, the precise sensitivity of IceCube-Gen2 to a neutrino signal from dark matter   annihilations in the Sun cannot yet be determined. But, given its larger volume and better angular resolution, an improvement of up to one order of magnitude could be expected with respect to the sensitivity of IceCube. For definiteness, we assume in our analysis that the sensitivity of IceCube-Gen2 is exactly a factor $10$ better than the current IceCube bounds.  

Important sensitivity improvements are also expected, within the next decade, in direct and indirect dark matter detection experiments. On the indirect detection front, the most significant one will likely come from CTA \cite{Mangano:2017tnd}, the Cherenkov Telescope Array. With more than 100 telescopes located in both hemispheres, CTA will play a crucial role in indirect (via gamma-rays) dark matter searches \cite{Conrad:2016jww}. A detailed analysis of the expected sensitivity of CTA to a dark matter signal (from the galactic center) was presented in \cite{Silverwood:2014yza}. In our figures, we use those results for the CTA sensitivity. On the direct detection front, new limits are expected, in the short term, from XENON1T and PANDAX-II, which are currently running. Farther into the future, new experiments such as XENON-nT or Darwin are expected to take lead and provide the most stringent limits.

\section{Results}
\label{sec:results}
%The neutrino signal from dark matter annihilations in the Sun depends, on the particle physics side, on just four parameters: the dark matter mass ($\mdm$), the dark matter scattering cross section off protons ($\sigma_p$), the dark matter annihilation branching fraction into different final states, and the dark matter total annihilation rate ($\sigma v$). 
When equilibrium is reached in the Sun between the capture and the annihilation rates, the limits on the neutrino signal can be displayed on the plane ($m_\chi$, $\sigma_p$) for a given final state. Since we do not want to assume equilibrium,  we need to find another way of displaying the four variables that determine the neutrino signal. A good choice seems to be the plane ($\sigma v$, $\sigma_p$) for a given final state and for representative values of the dark matter mass, say $10$ GeV, $100$ GeV, $1$ TeV, and $10$ TeV. In addition, since $\sigma_p$ could be due to spin-independent or spin-dependent interactions, it is useful to separately consider these two cases. That is the way the parameter space is displayed and analyzed in figures \ref{fig:sibb}-\ref{fig:sdtautau}.

This parameter space is constrained not only  by direct detection experiments and by neutrino detectors but also by indirect detection searches. The current limits from indirect detection are set by the Fermi-LAT observations of dwarf spheroidal galaxies, for $\mdm\leq 100$ GeV, and by the HESS limits from the galactic center at heavier masses. In figures \ref{fig:sibb}-\ref{fig:sdtautau}, the most stringent constraints on the parameter space are shown as well as the expected sensitivities to be reached in future experiments.
%The direct detection limits come instead from the XENON1T experiment for spin-independent interactions and from PICO-60 for spin-dependent ones. Both of them were released only recently. Neutrino detectors, on the other hand, can directly constrain the neutrino flux from dark matter annihilations in the Sun.  SuperKamiokande and IceCube currently provide the most stringent limits on such a flux, and important improvements are expected in the future from HyperKamiokande and IceCube-Gen2. 

\subsection{Capture via spin-independent interactions}
\label{sub:si}

\begin{figure}[tb]
\begin{center}
\includegraphics[scale=0.55]{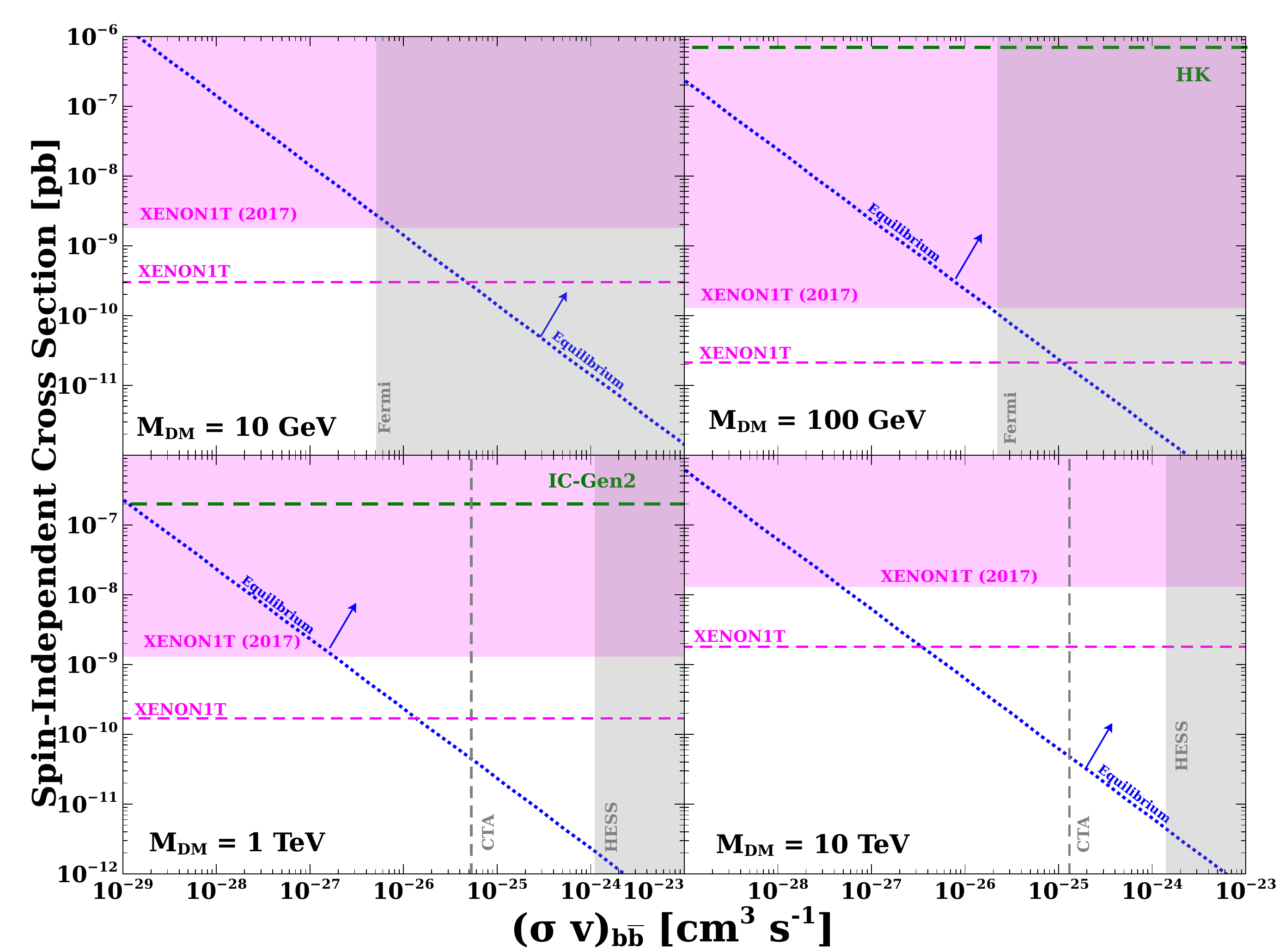}
\caption{The relevant parameter space for neutrino signals from dark matter annihilations in the Sun. In this figure it is assumed that the dark matter particle has only spin-independent interactions and that it annihilates into the $b\bar b$ final state.  Shaded areas and solid lines denote current bounds while future sensitivities are displayed as dashed lines. The dotted line is the boundary of the equilibrium region for the Sun.   \label{fig:sibb}} 
\end{center}
\end{figure}

\begin{figure}[tb]
\begin{center}
\includegraphics[scale=0.55]{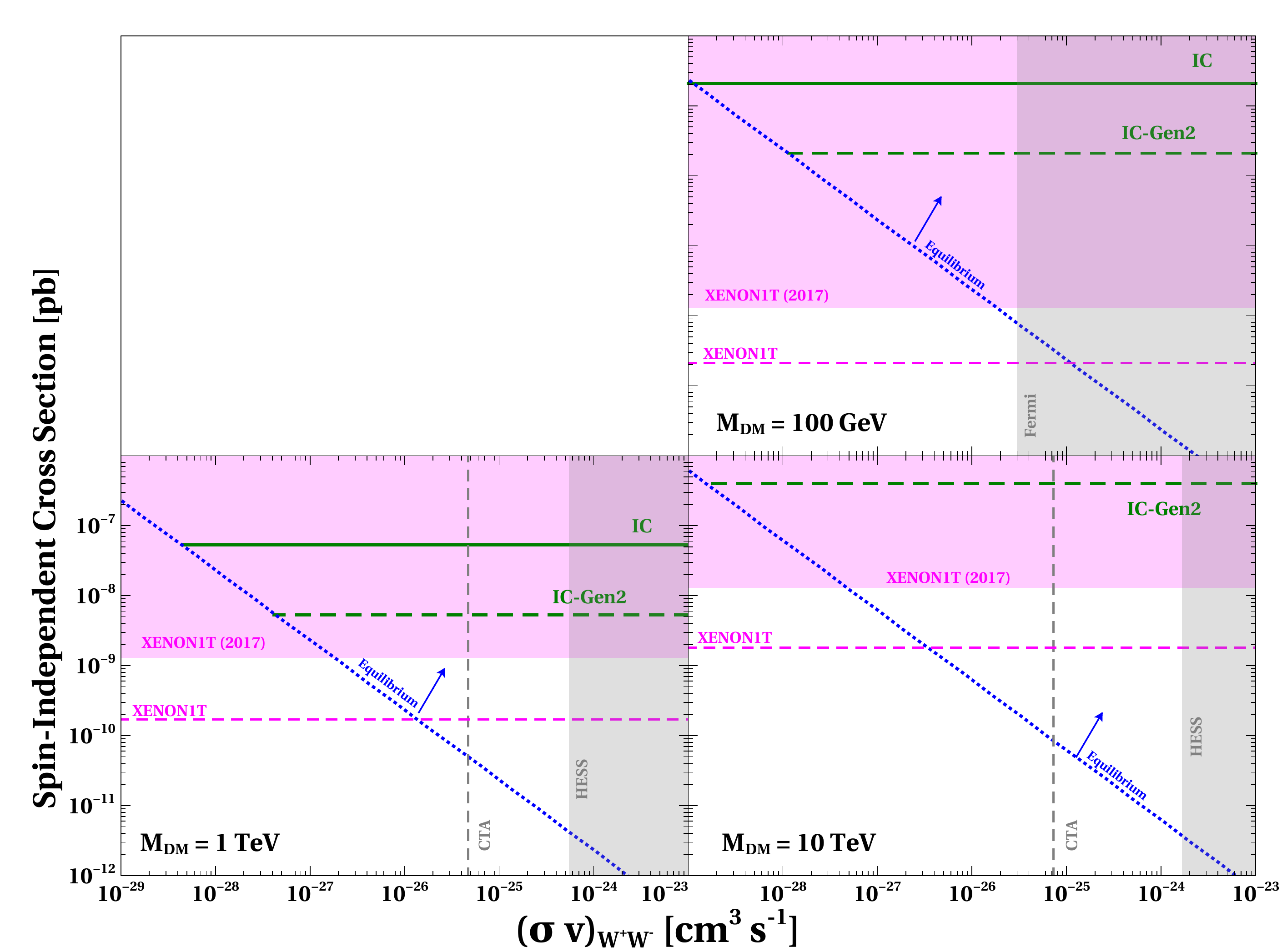}
\caption{The relevant parameter space for neutrino signals from dark matter annihilations in the Sun. In this figure it is assumed that the dark matter particle has only spin-independent interactions and that it annihilates into the $W^+W^-$ final state (the panel corresponding to $\mdm=10~$GeV is left empty because in that case the annihilation into $W^+W^-$ is not allowed).  Shaded areas and solid lines denote current bounds while future sensitivities are displayed as dashed lines. The dotted line is the boundary of the equilibrium region for the Sun. \label{fig:siww}} 
\end{center}
\end{figure}

\begin{figure}[tb]
\begin{center}
\includegraphics[scale=0.55]{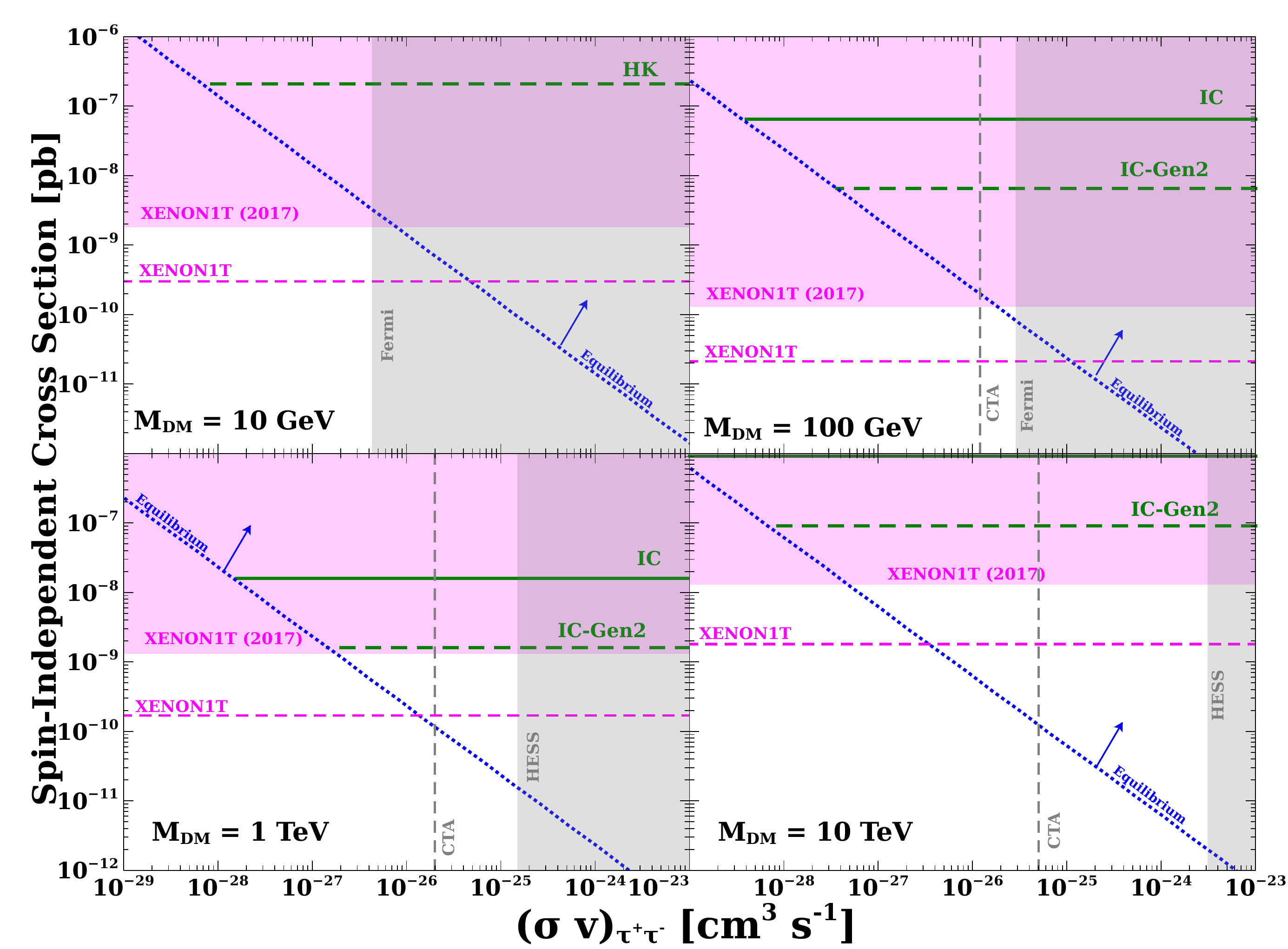}
\caption{The relevant parameter space for neutrino signals from dark matter annihilations in the Sun. In this figure it is assumed that the dark matter particle has only spin-independent interactions and that it  annihilates into the $\tau^+\tau^-$ final state.  Shaded areas and solid lines denote current bounds while future sensitivities are displayed as dashed lines. The dotted line is the boundary of the equilibrium region for the Sun.   \label{fig:sitautau}} 
\end{center}
\end{figure}

Our results for dark matter particles that have been captured in the Sun via  spin-independent interactions are shown, for the $b\bar b$, $W^+W^-$ and $\tau^+\tau^-$ annihilations channels, respectively in figures \ref{fig:sibb}-\ref{fig:sitautau}. Each figure consists of four panels corresponding to different dark matter masses: $10$ GeV, $100$ GeV, $1$ TeV, and $10$ TeV. In a given panel, the allowed regions of the parameter space, including current bounds (shaded areas and solid lines) and future prospects (dashed lines), are illustrated in the plane ($\sigma v$, $\sigma^{SI}_p$). In addition, the region where  equilibrium is reached between  capture and annihilation in the Sun   is shown as a dotted blue line.

For the $b\bar b$ final state, which produces a soft neutrino spectrum, we see from  figure \ref{fig:sibb} that future neutrino bounds, from HK and IC-Gen2, will only probe regions which are largely excluded by direct detection experiments. In fact, for dark matter masses of $10$ and $100$ GeV the entire equilibrium region is already ruled out by the bounds from Fermi-LAT and XENON1T, illustrating the complementarity between direct and indirect detection searches. For higher dark matter masses, the bounds on the dark matter scattering cross section and on the annihilation cross section weaken. The limits from direct detection experiments for dark matter masses much heavier than the target mass scale linearly with the dark matter mass since the local number density of dark matter particles is fixed. As for indirect dark matter detection, the limit weakens because the flux of gamma-rays from dark matter annihilation is inversely proportional to the dark matter mass squared. 
%The fewer gamma-rays the weaker are the bounds on the dark matter annihilation cross section. 
That is why, for higher dark matter masses the equilibrium regions is not entirely excluded, but it will be significantly reduced by the future bounds from XENON1T and CTA. 

For the $W^+W^-$ final state (figure \ref{fig:siww}), the neutrino constraints (and future sensitivities) are much stronger but still fall well inside the region excluded by direct detection experiments.  Notice that current bounds practically exclude the equilibrium region for $\mdm=100$~GeV and that future ones will do the same for $\mdm=1$~TeV. 

It is for the $\tau^+\tau^-$ final state that the neutrino detectors are most competitive, as illustrated in figure \ref{fig:sitautau}. In this case the future limits from IC-Gen2 will reach, for $\mdm=1$ TeV, almost the same sensitivity as current direct detection experiments, but they will still fail to  probe new viable regions of the parameter space.  

From figures \ref{fig:sibb}-\ref{fig:sitautau} we can conclude that if the dark matter has only spin-independent interactions, future experiments will be unable to detect a neutrino signal from dark matter annihilations in the Sun, irrespective of the dark matter mass or the annihilation final state. 

%\begin{figure}[tb]
%\begin{center}
%\includegraphics[scale=0.7]{SInunu}
%\caption{Results for the Spin-Independent case, $\nu\bar\nu$ channel. \label{fig:sinunu}} 
%\end{center}
%\end{figure} 

\subsection{Capture via spin-dependent interactions}
\label{sub:sd}

\begin{figure}[tb]
\begin{center}
\includegraphics[scale=0.55]{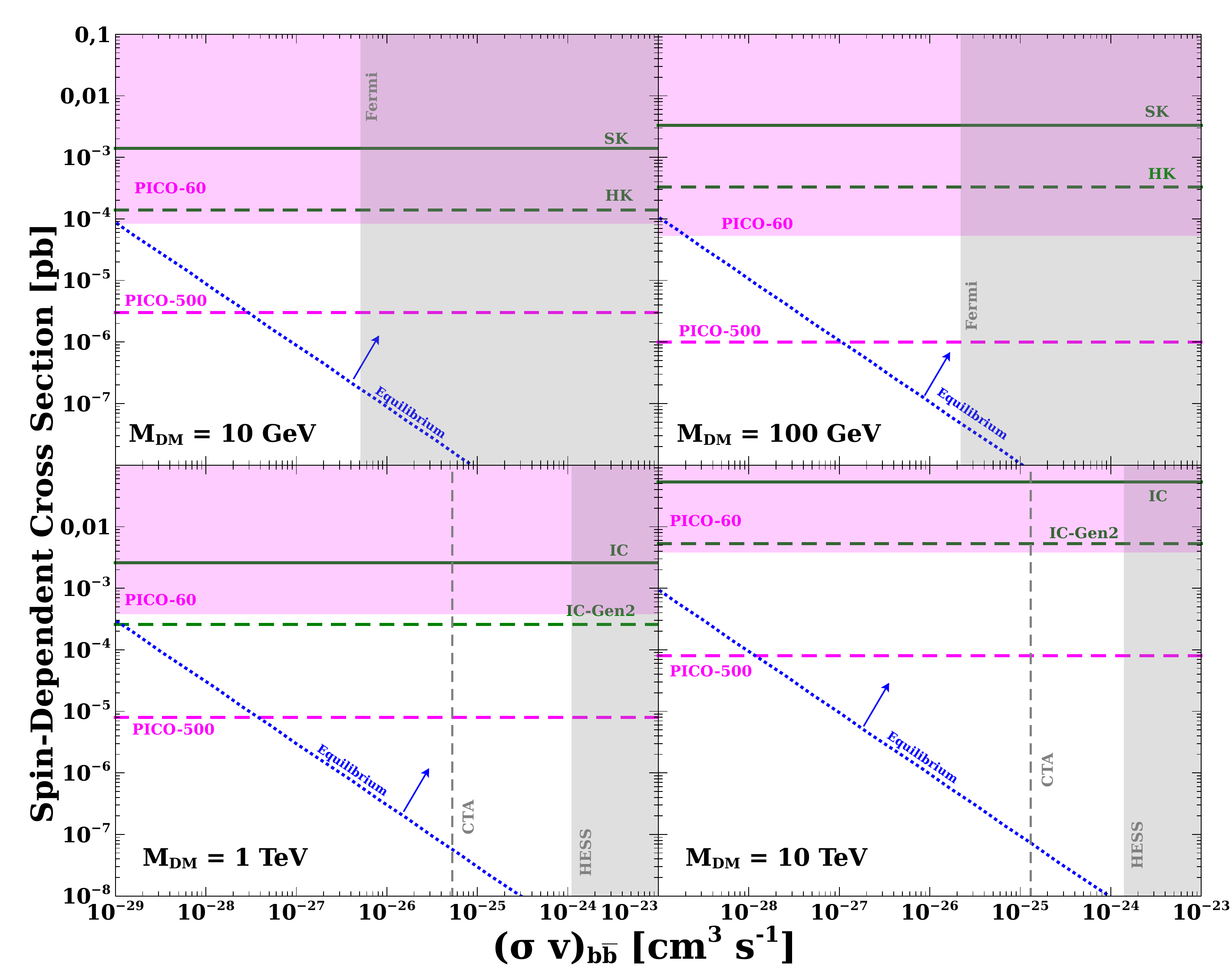}
\caption{The relevant parameter space for neutrino signals from dark matter annihilations in the Sun. In this figure it is assumed that the dark matter particle has only spin-dependent interactions and that it annihilates into the $b\bar b$ final state.  Shaded areas and solid lines denote current bounds while future sensitivities are displayed as dashed lines. The dotted line is the boundary of the equilibrium region for the Sun.   \label{fig:sdbb}} 
\end{center}
\end{figure} 

\begin{figure}[tb]
\begin{center}
\includegraphics[scale=0.55]{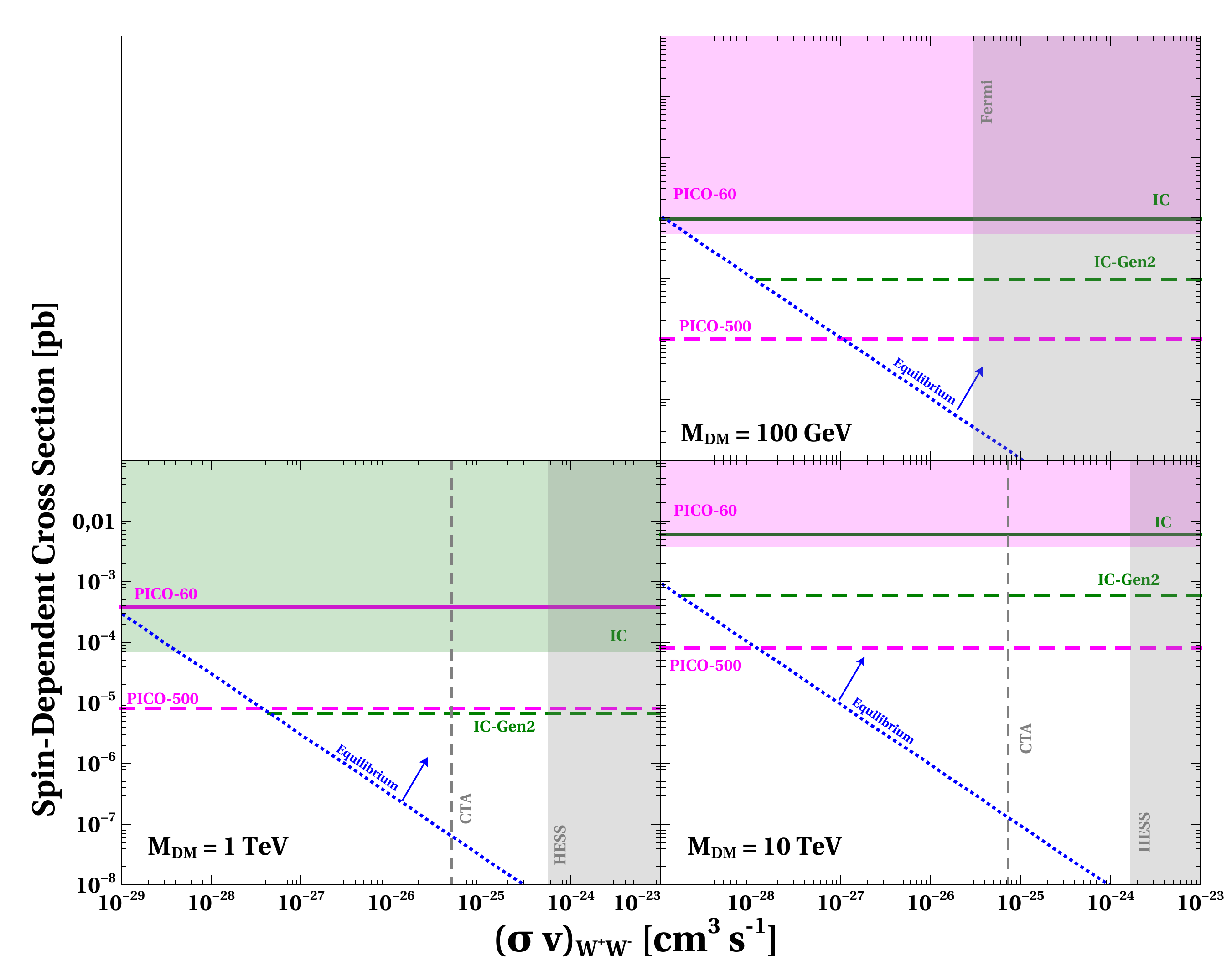}
\caption{The relevant parameter space for neutrino signals from dark matter annihilations in the Sun. In this figure it is assumed that the dark matter particle has only spin-dependent interactions and that it annihilates into the $W^+W^-$ final state.  Shaded areas and solid lines denote current bounds while future sensitivities are displayed as dashed lines. The dotted line is the boundary of the equilibrium region for the Sun.   \label{fig:sdww}} 
\end{center}
\end{figure}

\begin{figure}[tb]
\begin{center}
\includegraphics[scale=0.55]{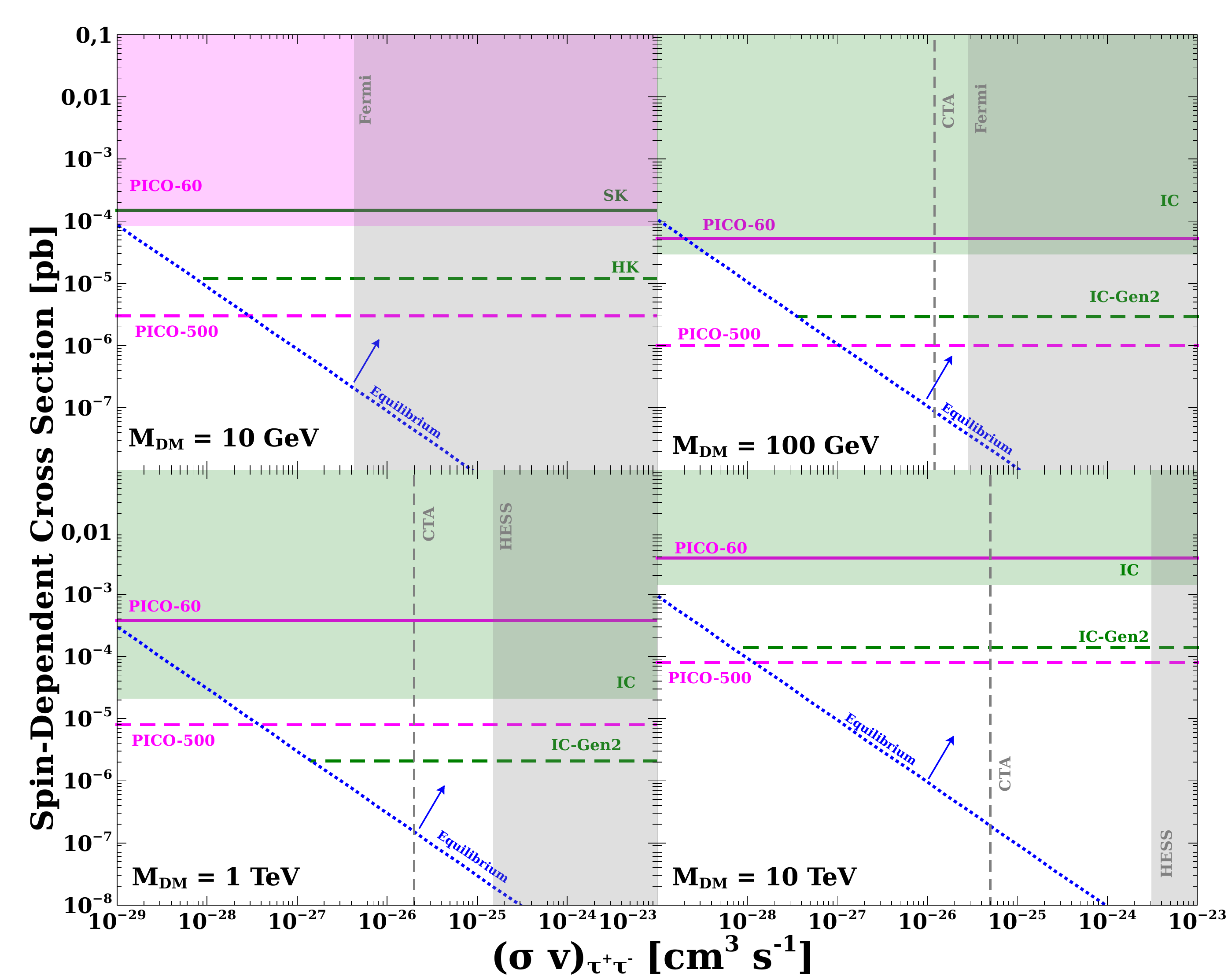}
\caption{The relevant parameter space for neutrino signals from dark matter annihilations in the Sun. In this figure it is assumed that the dark matter particle has only spin-dependent interactions and that it  annihilates into the $\tau^+\tau^-$ final state.  Shaded areas and solid lines denote current bounds while future sensitivities are displayed as dashed lines. The dotted line is the boundary of the equilibrium region for the Sun..  \label{fig:sdtautau}} 
\end{center}
\end{figure}

If the dark matter particle has spin-dependent interactions the detection prospects of a neutrino signal from the Sun improve significantly, as illustrated by figures \ref{fig:sdbb}-\ref{fig:sdtautau}. From Figure \ref{fig:sdbb}, where the final state $b\bar b$ is considered, it can be seen that, for all dark matter masses, current experiments probe only part of the equilibrium region, leaving plenty of room for a possible neutrino signal. It is only for $\mdm=1$~TeV, however, that future neutrino detectors will be able to probe a small region of the viable parameter space, and that region lies  well inside the expected sensitivity of PICO-500.

The final state $W^+W^-$ is analyzed in figure \ref{fig:sdww}. In this case we see that the bound on the spin-dependent cross section for $\mdm=1$~TeV is actually set by the IceCube experiment rather than by direct detection experiments. In fact,  for such a dark matter mass, the expected sensitivity of neutrino detectors and of dark matter detection experiments is similar.  That is, if  a signal were observed  in IceCube-Gen2, it should also be observed in PICO-500, and vice versa. For other dark matter masses, PICO-500 has a greater reach. In any case,  both experiments will probe new regions of the parameter space independently of the dark matter  mass. From the figure it can be seen that, over a wide region of the parameter space,   a neutrino signal from dark matter annihilations in the Sun could be observed in future neutrino experiments. If it were indeed observed, we would expect to observe a signal also in direct detection experiments.

Finally, we analyze the final state $\tau^+\tau^-$ in figure \ref{fig:sdtautau}. Notice that for $\mdm\geq 100$~GeV the current bounds on the spin-dependent cross section are set by IceCube.  In the future, both types of experiments, neutrino detectors and dark matter detection experiments, will probe  viable regions of the parameter space. For dark matter masses of order $1$~TeV, the reach of neutrino detectors is expected to be higher whereas at larger or smaller masses it is PICO-500 that has a higher reach. Thus, depending on the dark matter mass, it may be that only the neutrino signal is observed (for $\mdm\sim 1~\mathrm{TeV}$) or that both, the neutrino and the direct detection signals, are detected. Notice that at a dark matter mass of $10$~GeV it is Hyper-Kamiokande rather than IceCube-Gen2 that could detect the neutrino signal from dark matter annihilations in the Sun, illustrating the complementarity among different experiments.

%\begin{figure}[tb]
%\begin{center}
%\includegraphics[scale=0.7]{SDnunu}
%\caption{Results for the Spin-Dependent case, $\nu\bar\nu$ channel.  \label{fig:sdnunu}} 
%\end{center}
%\end{figure} 

\section{Discussion}

In our study we tried to be rather general: we did not consider a specific particle physics model,  but rather took the relevant properties of the dark matter particle as free parameters; we examine capture in the Sun via both spin-independent and spin-dependent interactions; for  dark matter annihilations,  the most  representative final states were separately considered; we varied the dark matter mass over  the entire  range of interest for a neutrino signal from the Sun; and, we did not assume equilibrium  between the capture and the annihilation rates in the Sun. Still, it is not difficult to envision even more generic frameworks where some of our assumptions do not hold.

For isospin-violating dark matter \cite{Feng:2011vu,Chen:2011vda,Zhou:2012qv,Kumar:2012uh}, for example,  the dark matter coupling to the proton and the neutron are different, so that $\sigma_p$ no longer suffices to characterize the dark matter-nucleon interactions. In that case, the signal regions will shift  for both, direct detection experiments and the neutrino flux  from the Sun, as recently emphasized in \cite{Yaguna:2016bga}. Regarding the dark matter interactions with nucleons, we assumed they were dominated by either spin-dependent or spin-independent operators, but more exotic possibilities, such as momentum dependent interactions or inelastic scattering, have also been considered in the literature \cite{Dedes:2009bk,Garani:2017jcj,Busoni:2017mhe,Blennow:2015hzp}. Such alternative scenarios lie beyond the scope of the present paper. 

%Models with suppressed couplings to quarks can be straightforwardly assessed by a rescaling if the capture rate is dominated by dark matter-nucleon scattering. Still within the leptophilic scenario, if  dark matter scattering cross section off nucleon features momentum dependence, the dark matter capture rate could be dominated by scattering off electrons, for dark matter masses below $1$~GeV \cite{Dedes:2009bk,Garani:2017jcj,Busoni:2017mhe}. For larger dark matter masses, as adopted in this work, the dark matter-nucleon scattering governs the capture rate validating our procedure even for leptophilic dark matter models.

%Moreover, we assumed that dark matter has no companion with nearly degenerate mass, so that inelastically scatter off nucleon is irrelevant \cite{Blennow:2015hzp}.

Our results, as presented in the previous section, are valid for a dark matter particle that annihilates entirely into a single final state --either $b\bar b$, $W^+W^-$ or $\tau^+\tau^-$. In certain models, such final states are indeed dominant and our conclusions directly apply to them,  whereas in other models more than one final state can contribute and one may need to \emph{interpolate} from the figures. In any case, this way of presenting the current bounds and the expected future sensitivities seems to be  the only way of keeping our results model independent, and is the same procedure followed by the experimental collaborations.   

\section{Conclusions}
We assessed the prospects for the detection of a neutrino signal from dark matter annihilations in the Sun taking into account current limits and expected future sensitivities from neutrino detectors as well as from  direct and indirect dark matter detection experiments. Throughout our analysis, we did not assume equilibrium between the capture and annihilation rates in the Sun but rather took the dark matter scattering and annihilation cross sections as free parameters. We considered capture via spin-independent and spin-dependent interactions and annihilations into three possible final states: $b\bar b$, $W^+W^-$, and $\tau^+\tau^-$, which give rise to both, soft and hard neutrino spectra. For spin-independent interactions, we found that current limits from direct detection experiments already preclude the observation of a neutrino signal in future experiments. For spin-dependent interactions, we did find regions of the parameter space where  future neutrino telescopes could play a role in the detection and the identification of the  dark matter particle.  Interestingly, most of such regions feature  a strong complementarity between future neutrino detectors and planned direct and indirect dark matter detection experiments, such as PICO-500 and CTA. In addition, we identified some regions of the parameter space that can be probed, via the neutrino flux from the Sun, only by future neutrino experiments. 
%by exploiting the interplay with ongoing and future direct and indirect detection experiments.

\acknowledgments
The authors would like to thank Manfred Lindner, Werner Rodejohann, Giorgio Arcadi and Joachim Kopp for discussions. CY is supported  by the DFG with grant RO 2516/5-1. The research of A.M. is supported by the ERC Advanced Grant No. 267985 (DaMeSyFla) and by the INFN. NF also acknowledges support from the research grant TAsP (Theoretical Astroparticle Physics) funded by the Istituto Nazionale di Fisica Nucleare (INFN). FSQ acknowledges support from MEC and ICTP-SAIFR FAPESP grant 2016/01343-7.

\bibliographystyle{JHEPfixed}
\bibliography{darkmatter}

\providecommand{\href}[2]{#2}\begingroup\raggedright\begin{thebibliography}{100}

\bibitem{Bertone:2004pz}
G.~Bertone, D.~Hooper, and J.~Silk, {\it {Particle dark matter: Evidence,
  candidates and constraints}},  {\em Phys. Rept.} {\bf 405} (2005) 279--390,
  [\href{http://xxx.lanl.gov/abs/hep-ph/0404175}{{\tt hep-ph/0404175}}].

\bibitem{Bertone:2016nfn}
G.~Bertone and D.~Hooper, {\it {A History of Dark Matter}},  {\em Submitted to:
  Rev. Mod. Phys.} (2016) [\href{http://xxx.lanl.gov/abs/1605.04909}{{\tt
  1605.04909}}].

\bibitem{Queiroz:2016sxf}
F.~S. Queiroz, W.~Rodejohann, and C.~E. Yaguna, {\it {Is the dark matter
  particle its own antiparticle?}},  {\em Phys. Rev.} {\bf D95} (2017), no.~9
  095010, [\href{http://xxx.lanl.gov/abs/1610.06581}{{\tt 1610.06581}}].

\bibitem{Kavanagh:2017hcl}
B.~J. Kavanagh, F.~S. Queiroz, W.~Rodejohann, and C.~E. Yaguna, {\it {Prospects
  for determining the particle/antiparticle nature of WIMP dark matter with
  direct detection experiments}},
  \href{http://xxx.lanl.gov/abs/1706.07819}{{\tt 1706.07819}}.

\bibitem{Pato:2010zk}
M.~Pato, L.~Baudis, G.~Bertone, R.~Ruiz~de Austri, L.~E. Strigari, and
  R.~Trotta, {\it {Complementarity of Dark Matter Direct Detection Targets}},
  {\em Phys. Rev.} {\bf D83} (2011) 083505,
  [\href{http://xxx.lanl.gov/abs/1012.3458}{{\tt 1012.3458}}].

\bibitem{Cirelli:2012tf}
M.~Cirelli, {\it {Indirect Searches for Dark Matter: a status review}},  {\em
  Pramana} {\bf 79} (2012) 1021--1043,
  [\href{http://xxx.lanl.gov/abs/1202.1454}{{\tt 1202.1454}}].

\bibitem{Cremonesi:2013bma}
E.~Del~Nobile, {\it {Halo-independent comparison of direct dark matter
  detection data: a review}},  {\em Adv. High Energy Phys.} {\bf 2014} (2014)
  604914, [\href{http://xxx.lanl.gov/abs/1404.4130}{{\tt 1404.4130}}].

\bibitem{Strigari:2013iaa}
L.~E. Strigari, {\it {Galactic Searches for Dark Matter}},  {\em Phys. Rept.}
  {\bf 531} (2013) 1--88, [\href{http://xxx.lanl.gov/abs/1211.7090}{{\tt
  1211.7090}}].

\bibitem{Iocco:2015xga}
F.~Iocco, M.~Pato, and G.~Bertone, {\it {Evidence for dark matter in the inner
  Milky Way}},  {\em Nature Phys.} {\bf 11} (2015) 245--248,
  [\href{http://xxx.lanl.gov/abs/1502.03821}{{\tt 1502.03821}}].

\bibitem{Conrad:2015bsa}
J.~Conrad, J.~Cohen-Tanugi, and L.~E. Strigari, {\it {WIMP searches with gamma
  rays in the Fermi era: challenges, methods and results}},  {\em J. Exp.
  Theor. Phys.} {\bf 121} (2015), no.~6 1104--1135,
  [\href{http://xxx.lanl.gov/abs/1503.06348}{{\tt 1503.06348}}]. [Zh. Eksp.
  Teor. Fiz.148,no.6,1257(2015)].

\bibitem{Mayet:2016zxu}
F.~Mayet {\em et.~al.}, {\it {A review of the discovery reach of directional
  Dark Matter detection}},  {\em Phys. Rept.} {\bf 627} (2016) 1--49,
  [\href{http://xxx.lanl.gov/abs/1602.03781}{{\tt 1602.03781}}].

\bibitem{Roszkowski:2017nbc}
L.~Roszkowski, E.~M. Sessolo, and S.~Trojanowski, {\it {WIMP dark matter
  candidates and searches - current issues and future prospects}},
  \href{http://xxx.lanl.gov/abs/1707.06277}{{\tt 1707.06277}}.

\bibitem{Battaglieri:2017aum}
M.~Battaglieri {\em et.~al.}, {\it {US Cosmic Visions: New Ideas in Dark Matter
  2017: Community Report}},  \href{http://xxx.lanl.gov/abs/1707.04591}{{\tt
  1707.04591}}.

\bibitem{Balazs:2017hxh}
C.~Balazs, J.~Conrad, B.~Farmer, T.~Jacques, T.~Li, M.~Meyer, F.~S. Queiroz,
  and M.~A. Sánchez-Conde, {\it {Sensitivity of the Cherenkov Telescope Array
  to the Detection of a Dark Matter Signal in comparison to Direct Detection
  and Collider Experiments}},  \href{http://xxx.lanl.gov/abs/1706.01505}{{\tt
  1706.01505}}.

\bibitem{Arcadi:2017kky}
G.~Arcadi, M.~Dutra, P.~Ghosh, M.~Lindner, Y.~Mambrini, M.~Pierre, S.~Profumo,
  and F.~S. Queiroz, {\it {The Waning of the WIMP? A Review of Models,
  Searches, and Constraints}},  \href{http://xxx.lanl.gov/abs/1703.07364}{{\tt
  1703.07364}}.

\bibitem{Acharya:2017ttl}
{\bf Cherenkov Telescope Array Consortium} Collaboration, B.~S. Acharya {\em
  et.~al.}, {\it {Science with the Cherenkov Telescope Array}},
  \href{http://xxx.lanl.gov/abs/1709.07997}{{\tt 1709.07997}}.

\bibitem{Angle:2011th}
{\bf XENON10} Collaboration, J.~Angle {\em et.~al.}, {\it {A search for light
  dark matter in XENON10 data}},  {\em Phys. Rev. Lett.} {\bf 107} (2011)
  051301, [\href{http://xxx.lanl.gov/abs/1104.3088}{{\tt 1104.3088}}].
  [Erratum: Phys. Rev. Lett.110,249901(2013)].

\bibitem{Aprile:2012nq}
{\bf XENON100 Collaboration} Collaboration, E.~Aprile {\em et.~al.}, {\it {Dark
  Matter Results from 225 Live Days of XENON100 Data}},  {\em Phys.Rev.Lett.}
  {\bf 109} (2012) 181301, [\href{http://xxx.lanl.gov/abs/1207.5988}{{\tt
  1207.5988}}].

\bibitem{Abe:2014zcd}
{\bf XMASS} Collaboration, K.~Abe {\em et.~al.}, {\it {Search for bosonic
  superweakly interacting massive dark matter particles with the XMASS-I
  detector}},  {\em Phys. Rev. Lett.} {\bf 113} (2014) 121301,
  [\href{http://xxx.lanl.gov/abs/1406.0502}{{\tt 1406.0502}}].

\bibitem{Angloher:2015ewa}
{\bf CRESST} Collaboration, G.~Angloher {\em et.~al.}, {\it {Results on light
  dark matter particles with a low-threshold CRESST-II detector}},  {\em Eur.
  Phys. J.} {\bf C76} (2016), no.~1 25,
  [\href{http://xxx.lanl.gov/abs/1509.01515}{{\tt 1509.01515}}].

\bibitem{Aprile:2016swn}
{\bf XENON100} Collaboration, E.~Aprile {\em et.~al.}, {\it {XENON100 Dark
  Matter Results from a Combination of 477 Live Days}},  {\em Phys. Rev.} {\bf
  D94} (2016), no.~12 122001, [\href{http://xxx.lanl.gov/abs/1609.06154}{{\tt
  1609.06154}}].

\bibitem{Cui:2017nnn}
{\bf PandaX-II} Collaboration, X.~Cui {\em et.~al.}, {\it {Dark Matter Results
  From 54-Ton-Day Exposure of PandaX-II Experiment}},
  \href{http://xxx.lanl.gov/abs/1708.06917}{{\tt 1708.06917}}.

\bibitem{Akerib:2017kat}
{\bf LUX} Collaboration, D.~S. Akerib {\em et.~al.}, {\it {Limits on
  spin-dependent WIMP-nucleon cross section obtained from the complete LUX
  exposure}},  {\em Phys. Rev. Lett.} {\bf 118} (2017), no.~25 251302,
  [\href{http://xxx.lanl.gov/abs/1705.03380}{{\tt 1705.03380}}].

\bibitem{Agnese:2017njq}
{\bf SuperCDMS} Collaboration, R.~Agnese {\em et.~al.}, {\it {Results from the
  Super Cryogenic Dark Matter Search (SuperCDMS) experiment at Soudan}},
  \href{http://xxx.lanl.gov/abs/1708.08869}{{\tt 1708.08869}}.

\bibitem{Aprile:2017yea}
{\bf XENON} Collaboration, E.~Aprile {\em et.~al.}, {\it {Search for Electronic
  Recoil Event Rate Modulation with 4 Years of XENON100 Data}},  {\em Phys.
  Rev. Lett.} {\bf 118} (2017), no.~10 101101,
  [\href{http://xxx.lanl.gov/abs/1701.00769}{{\tt 1701.00769}}].

\bibitem{Aguilar-Arevalo:2016zop}
{\bf DAMIC} Collaboration, A.~Aguilar-Arevalo {\em et.~al.}, {\it {First
  Direct-Detection Constraints on eV-Scale Hidden-Photon Dark Matter with DAMIC
  at SNOLAB}},  {\em Phys. Rev. Lett.} {\bf 118} (2017), no.~14 141803,
  [\href{http://xxx.lanl.gov/abs/1611.03066}{{\tt 1611.03066}}].

\bibitem{Essig:2017kqs}
R.~Essig, T.~Volansky, and T.-T. Yu, {\it {New Constraints and Prospects for
  sub-GeV Dark Matter Scattering off Electrons in Xenon}},  {\em Phys. Rev.}
  {\bf D96} (2017), no.~4 043017,
  [\href{http://xxx.lanl.gov/abs/1703.00910}{{\tt 1703.00910}}].

\bibitem{Cavoto:2017otc}
G.~Cavoto, F.~Luchetta, and A.~D. Polosa, {\it {Sub-GeV Dark Matter Detection
  with Electron Recoils in Carbon Nanotubes}},
  \href{http://xxx.lanl.gov/abs/1706.02487}{{\tt 1706.02487}}.

\bibitem{Davis:2017noy}
J.~H. Davis, {\it {Probing sub-GeV mass SIMP dark matter with a low-threshold
  surface experiment}},  \href{http://xxx.lanl.gov/abs/1708.01484}{{\tt
  1708.01484}}.

\bibitem{Undagoitia:2015gya}
T.~Marrodán~Undagoitia and L.~Rauch, {\it {Dark matter direct-detection
  experiments}},  {\em J. Phys.} {\bf G43} (2016), no.~1 013001,
  [\href{http://xxx.lanl.gov/abs/1509.08767}{{\tt 1509.08767}}].

\bibitem{Aprile:2017iyp}
{\bf XENON} Collaboration, E.~Aprile {\em et.~al.}, {\it {First Dark Matter
  Search Results from the XENON1T Experiment}},
  \href{http://xxx.lanl.gov/abs/1705.06655}{{\tt 1705.06655}}.

\bibitem{Aprile:2015uzo}
{\bf XENON} Collaboration, E.~Aprile {\em et.~al.}, {\it {Physics reach of the
  XENON1T dark matter experiment}},  {\em JCAP} {\bf 1604} (2016), no.~04 027,
  [\href{http://xxx.lanl.gov/abs/1512.07501}{{\tt 1512.07501}}].

\bibitem{Aalbers:2016jon}
{\bf DARWIN} Collaboration, J.~Aalbers {\em et.~al.}, {\it {DARWIN: towards the
  ultimate dark matter detector}},  {\em JCAP} {\bf 1611} (2016) 017,
  [\href{http://xxx.lanl.gov/abs/1606.07001}{{\tt 1606.07001}}].

\bibitem{Fatemighomi:2016ree}
{\bf DEAP-3600} Collaboration, N.~Fatemighomi, {\it {DEAP-3600 dark matter
  experiment}},  in {\em {35th International Symposium on Physics in Collision
  (PIC 2015) Coventry, United Kingdom, September 15-19, 2015}}, 2016.
\newblock \href{http://xxx.lanl.gov/abs/1609.07990}{{\tt 1609.07990}}.

\bibitem{Angloher:2014bua}
{\bf EURECA} Collaboration, G.~Angloher {\em et.~al.}, {\it {EURECA Conceptual
  Design Report}},  {\em Phys. Dark Univ.} {\bf 3} (2014) 41--74.

\bibitem{Amole:2017dex}
{\bf PICO} Collaboration, C.~Amole {\em et.~al.}, {\it {Dark Matter Search
  Results from the PICO-60 C$_3$F$_8$ Bubble Chamber}},  {\em Phys. Rev. Lett.}
  {\bf 118} (2017), no.~25 251301,
  [\href{http://xxx.lanl.gov/abs/1702.07666}{{\tt 1702.07666}}].

\bibitem{picoproject}
``Pico project.'' \url{http://www.picoexperiment.com}.
\newblock Accessed: 2010-09-30.

\bibitem{pico500A}
``Pico500.'' \url{http://www.picoexperiment.com/pico500.php}.
\newblock Accessed: 2010-09-30.

\bibitem{pico500}
``Pico bubble chambers, talk given by andrew sonnenschein at the cosmic visions
  workshop, university of maryland, march 23, 2017.''
  \url{https://indico.fnal.gov/getFile.py/access?contribId=93&sessionId=2&resId=0&materialId=slides&confId=13702}.
\newblock Accessed: 2010-09-30.

\bibitem{Colafrancesco:2006he}
S.~Colafrancesco, S.~Profumo, and P.~Ullio, {\it {Detecting dark matter WIMPs
  in the Draco dwarf: A multi-wavelength perspective}},  {\em Phys. Rev.} {\bf
  D75} (2007) 023513, [\href{http://xxx.lanl.gov/abs/astro-ph/0607073}{{\tt
  astro-ph/0607073}}].

\bibitem{Abramowski:2011hc}
{\bf HESS} Collaboration, A.~Abramowski {\em et.~al.}, {\it {Search for a Dark
  Matter annihilation signal from the Galactic Center halo with H.E.S.S}},
  {\em Phys. Rev. Lett.} {\bf 106} (2011) 161301,
  [\href{http://xxx.lanl.gov/abs/1103.3266}{{\tt 1103.3266}}].

\bibitem{Hooper:2012sr}
D.~Hooper, C.~Kelso, and F.~S. Queiroz, {\it {Stringent and Robust Constraints
  on the Dark Matter Annihilation Cross Section From the Region of the Galactic
  Center}},  {\em Astropart. Phys.} {\bf 46} (2013) 55--70,
  [\href{http://xxx.lanl.gov/abs/1209.3015}{{\tt 1209.3015}}].

\bibitem{Bringmann:2012ez}
T.~Bringmann and C.~Weniger, {\it {Gamma Ray Signals from Dark Matter:
  Concepts, Status and Prospects}},  {\em Phys. Dark Univ.} {\bf 1} (2012)
  194--217, [\href{http://xxx.lanl.gov/abs/1208.5481}{{\tt 1208.5481}}].

\bibitem{Baratella:2013fya}
P.~Baratella, M.~Cirelli, A.~Hektor, J.~Pata, M.~Piibeleht, and A.~Strumia,
  {\it {PPPC 4 DM$\nu$: a Poor Particle Physicist Cookbook for Neutrinos from
  Dark Matter annihilations in the Sun}},  {\em JCAP} {\bf 1403} (2014) 053,
  [\href{http://xxx.lanl.gov/abs/1312.6408}{{\tt 1312.6408}}].

\bibitem{Giesen:2015ufa}
G.~Giesen, M.~Boudaud, Y.~Génolini, V.~Poulin, M.~Cirelli, P.~Salati, and
  P.~D. Serpico, {\it {AMS-02 antiprotons, at last! Secondary astrophysical
  component and immediate implications for Dark Matter}},  {\em JCAP} {\bf
  1509} (2015), no.~09 023, [\href{http://xxx.lanl.gov/abs/1504.04276}{{\tt
  1504.04276}}].

\bibitem{Gonzalez-Morales:2017jkx}
A.~X. Gonzalez-Morales, S.~Profumo, and J.~Reynoso-Córdova, {\it {Prospects
  for indirect MeV Dark Matter detection with Gamma Rays in light of Cosmic
  Microwave Background Constraints}},
  \href{http://xxx.lanl.gov/abs/1705.00777}{{\tt 1705.00777}}.

\bibitem{Ackermann:2015zua}
{\bf Fermi-LAT} Collaboration, M.~Ackermann {\em et.~al.}, {\it {Searching for
  Dark Matter Annihilation from Milky Way Dwarf Spheroidal Galaxies with Six
  Years of Fermi Large Area Telescope Data}},  {\em Phys. Rev. Lett.} {\bf 115}
  (2015), no.~23 231301, [\href{http://xxx.lanl.gov/abs/1503.02641}{{\tt
  1503.02641}}].

\bibitem{Drlica-Wagner:2015xua}
{\bf DES, Fermi-LAT} Collaboration, A.~Drlica-Wagner {\em et.~al.}, {\it
  {Search for Gamma-Ray Emission from DES Dwarf Spheroidal Galaxy Candidates
  with Fermi-LAT Data}},  {\em Astrophys. J.} {\bf 809} (2015), no.~1 L4,
  [\href{http://xxx.lanl.gov/abs/1503.02632}{{\tt 1503.02632}}].

\bibitem{Abdallah:2016ygi}
{\bf H.E.S.S.} Collaboration, H.~Abdallah {\em et.~al.}, {\it {Search for dark
  matter annihilations towards the inner Galactic halo from 10 years of
  observations with H.E.S.S}},  {\em Phys. Rev. Lett.} {\bf 117} (2016), no.~11
  111301, [\href{http://xxx.lanl.gov/abs/1607.08142}{{\tt 1607.08142}}].

\bibitem{Acharya:2013sxa}
{\bf CTA Consortium} Collaboration, B.~S. Acharya {\em et.~al.}, {\it
  {Introducing the CTA concept}},  {\em Astropart. Phys.} {\bf 43} (2013)
  3--18.

\bibitem{Silverwood:2014yza}
H.~Silverwood, C.~Weniger, P.~Scott, and G.~Bertone, {\it {A realistic
  assessment of the CTA sensitivity to dark matter annihilation}},  {\em JCAP}
  {\bf 1503} (2015), no.~03 055, [\href{http://xxx.lanl.gov/abs/1408.4131}{{\tt
  1408.4131}}].

\bibitem{Hutten:2016jko}
M.~Hütten, C.~Combet, G.~Maier, and D.~Maurin, {\it {Dark matter substructure
  modelling and sensitivity of the Cherenkov Telescope Array to Galactic dark
  halos}},  {\em JCAP} {\bf 1609} (2016), no.~09 047,
  [\href{http://xxx.lanl.gov/abs/1606.04898}{{\tt 1606.04898}}].

\bibitem{Roszkowski:2016bhs}
L.~Roszkowski, E.~M. Sessolo, S.~Trojanowski, and A.~J. Williams, {\it
  {Reconstructing WIMP properties through an interplay of signal measurements
  in direct detection, Fermi-LAT, and CTA searches for dark matter}},
  \href{http://xxx.lanl.gov/abs/1603.06519}{{\tt 1603.06519}}.

\bibitem{Conrad:2016jww}
J.~Conrad, {\it {CTA in the Context of Searches for Particle Dark Matter - a
  glimpse}},  {\em AIP Conf. Proc.} {\bf 1792} (2017), no.~1 030002,
  [\href{http://xxx.lanl.gov/abs/1610.03258}{{\tt 1610.03258}}].

\bibitem{Morselli:2017ree}
{\bf for the CTA Consortium} Collaboration, A.~Morselli, {\it {The Dark Matter
  Programme of the Cherenkov Telescope Array}},  {\em PoS} {\bf ICRC2017}
  (2017) 921, [\href{http://xxx.lanl.gov/abs/1709.01483}{{\tt 1709.01483}}].

\bibitem{Choi:2015ara}
{\bf Super-Kamiokande} Collaboration, K.~Choi {\em et.~al.}, {\it {Search for
  neutrinos from annihilation of captured low-mass dark matter particles in the
  Sun by Super-Kamiokande}},  {\em Phys. Rev. Lett.} {\bf 114} (2015), no.~14
  141301, [\href{http://xxx.lanl.gov/abs/1503.04858}{{\tt 1503.04858}}].

\bibitem{Aartsen:2016zhm}
{\bf IceCube} Collaboration, M.~G. Aartsen {\em et.~al.}, {\it {Search for
  annihilating dark matter in the Sun with 3 years of IceCube data}},  {\em
  Eur. Phys. J.} {\bf C77} (2017), no.~3 146,
  [\href{http://xxx.lanl.gov/abs/1612.05949}{{\tt 1612.05949}}].

\bibitem{ElAisati:2017ppn}
C.~El~Aisati, C.~Garcia-Cely, T.~Hambye, and L.~Vanderheyden, {\it {Prospects
  for discovering a neutrino line induced by dark matter annihilation}},
  \href{http://xxx.lanl.gov/abs/1706.06600}{{\tt 1706.06600}}.

\bibitem{Ardid:2017lry}
M.~Ardid, I.~Felis, A.~Herrero, and J.~A. Martínez-Mora, {\it {Constraining
  Secluded Dark Matter models with the public data from the 79-string IceCube
  search for dark matter in the Sun}},  {\em JCAP} {\bf 1704} (2017), no.~04
  010, [\href{http://xxx.lanl.gov/abs/1701.08863}{{\tt 1701.08863}}].

\bibitem{Queiroz:2016zwd}
F.~S. Queiroz, C.~E. Yaguna, and C.~Weniger, {\it {Gamma-ray Limits on Neutrino
  Lines}},  {\em JCAP} {\bf 1605} (2016), no.~05 050,
  [\href{http://xxx.lanl.gov/abs/1602.05966}{{\tt 1602.05966}}].

\bibitem{Esmaili:2009ks}
A.~Esmaili and Y.~Farzan, {\it {On the Oscillation of Neutrinos Produced by the
  Annihilation of Dark Matter inside the Sun}},  {\em Phys. Rev.} {\bf D81}
  (2010) 113010, [\href{http://xxx.lanl.gov/abs/0912.4033}{{\tt 0912.4033}}].

\bibitem{Taoso:2010tg}
M.~Taoso, F.~Iocco, G.~Meynet, G.~Bertone, and P.~Eggenberger, {\it {Effect of
  low mass dark matter particles on the Sun}},  {\em Phys. Rev.} {\bf D82}
  (2010) 083509, [\href{http://xxx.lanl.gov/abs/1005.5711}{{\tt 1005.5711}}].

\bibitem{Fukushima:2012sp}
K.~Fukushima, Y.~Gao, J.~Kumar, and D.~Marfatia, {\it {Bremsstrahlung
  signatures of dark matter annihilation in the Sun}},  {\em Phys. Rev.} {\bf
  D86} (2012) 076014, [\href{http://xxx.lanl.gov/abs/1208.1010}{{\tt
  1208.1010}}].

\bibitem{Kundu:2011ek}
S.~Kundu and P.~Bhattacharjee, {\it {Neutrinos from WIMP annihilation in the
  Sun : Implications of a self-consistent model of the Milky Way's dark matter
  halo}},  {\em Phys. Rev.} {\bf D85} (2012) 123533,
  [\href{http://xxx.lanl.gov/abs/1106.5711}{{\tt 1106.5711}}].

\bibitem{Gao:2011bq}
Y.~Gao, J.~Kumar, and D.~Marfatia, {\it {Isospin-Violating Dark Matter in the
  Sun}},  {\em Phys. Lett.} {\bf B704} (2011) 534--540,
  [\href{http://xxx.lanl.gov/abs/1108.0518}{{\tt 1108.0518}}].

\bibitem{Ibarra:2013eba}
A.~Ibarra, M.~Totzauer, and S.~Wild, {\it {High-energy neutrino signals from
  the Sun in dark matter scenarios with internal bremsstrahlung}},  {\em JCAP}
  {\bf 1312} (2013) 043, [\href{http://xxx.lanl.gov/abs/1311.1418}{{\tt
  1311.1418}}].

\bibitem{Choi:2013eda}
K.~Choi, C.~Rott, and Y.~Itow, {\it {Impact of the dark matter velocity
  distribution on capture rates in the Sun}},  {\em JCAP} {\bf 1405} (2014)
  049, [\href{http://xxx.lanl.gov/abs/1312.0273}{{\tt 1312.0273}}].

\bibitem{Chen:2014oaa}
C.-S. Chen, F.-F. Lee, G.-L. Lin, and Y.-H. Lin, {\it {Probing Dark Matter
  Self-Interaction in the Sun with IceCube-PINGU}},  {\em JCAP} {\bf 1410}
  (2014), no.~10 049, [\href{http://xxx.lanl.gov/abs/1408.5471}{{\tt
  1408.5471}}].

\bibitem{Berger:2014sqa}
J.~Berger, Y.~Cui, and Y.~Zhao, {\it {Detecting Boosted Dark Matter from the
  Sun with Large Volume Neutrino Detectors}},  {\em JCAP} {\bf 1502} (2015),
  no.~02 005, [\href{http://xxx.lanl.gov/abs/1410.2246}{{\tt 1410.2246}}].

\bibitem{Blumenthal:2014cwa}
J.~Blumenthal, P.~Gretskov, M.~Krämer, and C.~Wiebusch, {\it {Effective field
  theory interpretation of searches for dark matter annihilation in the Sun
  with the IceCube Neutrino Observatory}},  {\em Phys. Rev.} {\bf D91} (2015),
  no.~3 035002, [\href{http://xxx.lanl.gov/abs/1411.5917}{{\tt 1411.5917}}].

\bibitem{Catena:2015uha}
R.~Catena and B.~Schwabe, {\it {Form factors for dark matter capture by the Sun
  in effective theories}},  {\em JCAP} {\bf 1504} (2015), no.~04 042,
  [\href{http://xxx.lanl.gov/abs/1501.03729}{{\tt 1501.03729}}].

\bibitem{Kouvaris:2015nsa}
C.~Kouvaris, {\it {Probing Light Dark Matter via Evaporation from the Sun}},
  {\em Phys. Rev.} {\bf D92} (2015), no.~7 075001,
  [\href{http://xxx.lanl.gov/abs/1506.04316}{{\tt 1506.04316}}].

\bibitem{Rott:2016mzs}
C.~Rott, S.~In, J.~Kumar, and D.~Yaylali, {\it {Directional Searches at DUNE
  for Sub-GeV Monoenergetic Neutrinos Arising from Dark Matter Annihilation in
  the Sun}},  {\em JCAP} {\bf 1701} (2017), no.~01 016,
  [\href{http://xxx.lanl.gov/abs/1609.04876}{{\tt 1609.04876}}].

\bibitem{Feng:2016ijc}
J.~L. Feng, J.~Smolinsky, and P.~Tanedo, {\it {Detecting dark matter through
  dark photons from the Sun: Charged particle signatures}},  {\em Phys. Rev.}
  {\bf D93} (2016), no.~11 115036,
  [\href{http://xxx.lanl.gov/abs/1602.01465}{{\tt 1602.01465}}].

\bibitem{Abe:2011ts}
K.~Abe {\em et.~al.}, {\it {Letter of Intent: The Hyper-Kamiokande Experiment
  --- Detector Design and Physics Potential ---}},
  \href{http://xxx.lanl.gov/abs/1109.3262}{{\tt 1109.3262}}.

\bibitem{Aartsen:2014njl}
{\bf IceCube} Collaboration, M.~G. Aartsen {\em et.~al.}, {\it {IceCube-Gen2: A
  Vision for the Future of Neutrino Astronomy in Antarctica}},
  \href{http://xxx.lanl.gov/abs/1412.5106}{{\tt 1412.5106}}.

\bibitem{Katz:2006wv}
U.~F. Katz, {\it {KM3NeT: Towards a km**3 Mediterranean Neutrino Telescope}},
  {\em Nucl. Instrum. Meth.} {\bf A567} (2006) 457--461,
  [\href{http://xxx.lanl.gov/abs/astro-ph/0606068}{{\tt astro-ph/0606068}}].

\bibitem{Adrian-Martinez:2016fdl}
{\bf KM3Net} Collaboration, S.~Adrian-Martinez {\em et.~al.}, {\it {Letter of
  intent for KM3NeT 2.0}},  {\em J. Phys.} {\bf G43} (2016), no.~8 084001,
  [\href{http://xxx.lanl.gov/abs/1601.07459}{{\tt 1601.07459}}].

\bibitem{Gondolo:2004sc}
P.~Gondolo, J.~Edsjo, P.~Ullio, L.~Bergstrom, M.~Schelke, and E.~A. Baltz, {\it
  {DarkSUSY: Computing supersymmetric dark matter properties numerically}},
  {\em JCAP} {\bf 0407} (2004) 008,
  [\href{http://xxx.lanl.gov/abs/astro-ph/0406204}{{\tt astro-ph/0406204}}].

\bibitem{Belanger:2013oya}
G.~Belanger, F.~Boudjema, A.~Pukhov, and A.~Semenov, {\it {micrOMEGAs3: A
  program for calculating dark matter observables}},  {\em Comput. Phys.
  Commun.} {\bf 185} (2014) 960--985,
  [\href{http://xxx.lanl.gov/abs/1305.0237}{{\tt 1305.0237}}].

\bibitem{Cirelli:2005gh}
M.~Cirelli, N.~Fornengo, T.~Montaruli, I.~A. Sokalski, A.~Strumia, and
  F.~Vissani, {\it {Spectra of neutrinos from dark matter annihilations}},
  {\em Nucl. Phys.} {\bf B727} (2005) 99--138,
  [\href{http://xxx.lanl.gov/abs/hep-ph/0506298}{{\tt hep-ph/0506298}}].
  [Erratum: Nucl. Phys.B790,338(2008)].

\bibitem{Boliev:2013ai}
M.~M. Boliev, S.~V. Demidov, S.~P. Mikheyev, and O.~V. Suvorova, {\it {Search
  for muon signal from dark matter annihilations inthe Sun with the Baksan
  Underground Scintillator Telescope for 24.12 years}},  {\em JCAP} {\bf 1309}
  (2013) 019, [\href{http://xxx.lanl.gov/abs/1301.1138}{{\tt 1301.1138}}].

\bibitem{Adrian-Martinez:2016gti}
{\bf ANTARES} Collaboration, S.~Adrian-Martinez {\em et.~al.}, {\it {Limits on
  Dark Matter Annihilation in the Sun using the ANTARES Neutrino Telescope}},
  {\em Phys. Lett.} {\bf B759} (2016) 69--74,
  [\href{http://xxx.lanl.gov/abs/1603.02228}{{\tt 1603.02228}}].

\bibitem{Akerib:2016vxi}
{\bf LUX} Collaboration, D.~S. Akerib {\em et.~al.}, {\it {Results from a
  search for dark matter in the complete LUX exposure}},  {\em Phys. Rev.
  Lett.} {\bf 118} (2017), no.~2 021303,
  [\href{http://xxx.lanl.gov/abs/1608.07648}{{\tt 1608.07648}}].

\bibitem{Abe:2015zbg}
{\bf Hyper-Kamiokande Proto-Collaboration} Collaboration, K.~Abe {\em et.~al.},
  {\it {Physics potential of a long-baseline neutrino oscillation experiment
  using a J-PARC neutrino beam and Hyper-Kamiokande}},  {\em PTEP} {\bf 2015}
  (2015) 053C02, [\href{http://xxx.lanl.gov/abs/1502.05199}{{\tt 1502.05199}}].

\bibitem{Yano:2016rkf}
{\bf Hyper-Kamiokande proto} Collaboration, T.~Yano, {\it {Neutrino
  astrophysics with Hyper-Kamiokande}},  {\em J. Phys. Conf. Ser.} {\bf 718}
  (2016), no.~6 062071.

\bibitem{Yokoyama:2017mnt}
{\bf Hyper-Kamiokande Proto} Collaboration, M.~Yokoyama, {\it {The
  Hyper-Kamiokande Experiment}},  in {\em {Prospects in Neutrino Physics
  (NuPhys2016) London, London, United Kingdom, December 12-14, 2016}}, 2017.
\newblock \href{http://xxx.lanl.gov/abs/1705.00306}{{\tt 1705.00306}}.

\bibitem{Labarga:2017rfo}
{\bf Hyper-Kamiokande proto} Collaboration, L.~Labarga, {\it {Astrophysics
  Potential of Hyper-Kamiokande}},  {\em PoS} {\bf ICHEP2016} (2017) 828.

\bibitem{Blaufuss:2015muc}
{\bf IceCube-Gen2} Collaboration, E.~Blaufuss, C.~Kopper, and C.~Haack, {\it
  {The IceCube-Gen2 High Energy Array}},  {\em PoS} {\bf ICRC2015} (2016) 1146.

\bibitem{Meagher:2017zvi}
{\bf IceCube} Collaboration, K.~J. Meagher, {\it {Neutrino Astronomy with
  IceCube and Beyond}},  in {\em {Prospects in Neutrino Physics (NuPhys2016)
  London, London, United Kingdom, December 12-14, 2016}}, 2017.
\newblock \href{http://xxx.lanl.gov/abs/1705.00383}{{\tt 1705.00383}}.

\bibitem{Mangano:2017tnd}
{\bf CTA Consortium} Collaboration, S.~Mangano, {\it {Cherenkov Telescope Array
  Status Report}},  \href{http://xxx.lanl.gov/abs/1705.07805}{{\tt
  1705.07805}}.

\bibitem{Feng:2011vu}
J.~L. Feng, J.~Kumar, D.~Marfatia, and D.~Sanford, {\it {Isospin-Violating Dark
  Matter}},  {\em Phys. Lett.} {\bf B703} (2011) 124--127,
  [\href{http://xxx.lanl.gov/abs/1102.4331}{{\tt 1102.4331}}].

\bibitem{Chen:2011vda}
S.-L. Chen and Y.~Zhang, {\it {Isospin-Violating Dark Matter and Neutrinos From
  the Sun}},  {\em Phys. Rev.} {\bf D84} (2011) 031301,
  [\href{http://xxx.lanl.gov/abs/1106.4044}{{\tt 1106.4044}}].

\bibitem{Zhou:2012qv}
Y.-F. Zhou, {\it {Direct and indirect constraints on isospin-violating dark
  matter}},  {\em Nucl. Phys. Proc. Suppl.} {\bf 246-247} (2014) 99--105,
  [\href{http://xxx.lanl.gov/abs/1212.2043}{{\tt 1212.2043}}].

\bibitem{Kumar:2012uh}
J.~Kumar, J.~G. Learned, S.~Smith, and K.~Richardson, {\it {Tools for Studying
  Low-Mass Dark Matter at Neutrino Detectors}},  {\em Phys. Rev.} {\bf D86}
  (2012) 073002, [\href{http://xxx.lanl.gov/abs/1204.5120}{{\tt 1204.5120}}].

\bibitem{Yaguna:2016bga}
C.~E. Yaguna, {\it {Isospin-violating dark matter in the light of recent
  data}},  {\em Phys. Rev.} {\bf D95} (2017), no.~5 055015,
  [\href{http://xxx.lanl.gov/abs/1610.08683}{{\tt 1610.08683}}].

\bibitem{Dedes:2009bk}
A.~Dedes, I.~Giomataris, K.~Suxho, and J.~D. Vergados, {\it {Searching for
  Secluded Dark Matter via Direct Detection of Recoiling Nuclei as well as Low
  Energy Electrons}},  {\em Nucl. Phys.} {\bf B826} (2010) 148--173,
  [\href{http://xxx.lanl.gov/abs/0907.0758}{{\tt 0907.0758}}].

\bibitem{Garani:2017jcj}
R.~Garani and S.~Palomares-Ruiz, {\it {Dark matter in the Sun: scattering off
  electrons vs nucleons}},  {\em JCAP} {\bf 1705} (2017), no.~05 007,
  [\href{http://xxx.lanl.gov/abs/1702.02768}{{\tt 1702.02768}}].

\bibitem{Busoni:2017mhe}
G.~Busoni, A.~De~Simone, P.~Scott, and A.~C. Vincent, {\it {Evaporation and
  scattering of momentum- and velocity-dependent dark matter in the Sun}},
  \href{http://xxx.lanl.gov/abs/1703.07784}{{\tt 1703.07784}}.

\bibitem{Blennow:2015hzp}
M.~Blennow, S.~Clementz, and J.~Herrero-Garcia, {\it {Pinning down inelastic
  dark matter in the Sun and in direct detection}},  {\em JCAP} {\bf 1604}
  (2016), no.~04 004, [\href{http://xxx.lanl.gov/abs/1512.03317}{{\tt
  1512.03317}}].

\end{thebibliography}\endgroup

\end{document}